\documentclass{article}
\usepackage[OT1]{fontenc}  
\usepackage{spconf}
\usepackage{amssymb,amsmath,bm,graphicx,booktabs}
\usepackage{fancyhdr}
\fancypagestyle{copyright}{%
  \fancyhf{}
  
  \fancyfoot[L]{\ninept \copyright\ 2022 IEEE.  Personal use of this material is permitted.  Permission from IEEE must be obtained for all other uses, in any current or future media, including reprinting/republishing this material for advertising or promotional purposes, creating new collective works, for resale or redistribution to servers or lists, or reuse of any copyrighted component of this work in other works.}
}

\newcommand{\N}{\mathbb{N}}

\newcommand{\R}{\mathbb{R}}

\newcommand{\parens}[1]{\left( #1 \right)}    
\newcommand{\Verts}[1]{\left\Vert #1 \right\Vert}  

\title{AN INVESTIGATION OF THE EFFECTIVENESS OF PHASE FOR AUDIO CLASSIFICATION}

\name{Shunsuke Hidaka$^1$, Kohei Wakamiya$^2$, Tokihiko Kaburagi$^2$}
\address{
  $^1$Graduate School of Design, Kyushu University, Japan,
  $^2$Faculty of Design, Kyushu University, Japan}

\begin{document}
\ninept  
\maketitle
\thispagestyle{copyright}

\begin{abstract}
\vspace{-.5\baselineskip}
While log-amplitude mel-spectrogram has widely been used as the feature representation for processing speech based on deep learning,
the effectiveness of another aspect of speech spectrum, i.e., phase information, was shown recently for tasks such as speech enhancement and source separation.
In this study, we extensively investigated the effectiveness of including phase information of signals for eight audio classification tasks.
We constructed a learnable front-end that can compute the phase and its derivatives based on a time-frequency representation with mel-like frequency axis.
As a result, experimental results showed significant performance improvement for musical pitch detection, musical instrument detection, language identification, speaker identification, and birdsong detection.
On the other hand, overfitting to the recording condition was observed for some tasks when the instantaneous frequency was used.
The results implied that the relationship between the phase values of adjacent elements is more important than the phase itself in audio classification.
\end{abstract}

\begin{keywords}
Time-frequency representation,
group delay,
instantaneous frequency,
learnable filterbank,
deep learning
\end{keywords}

\section{INTRODUCTION}
\label{sec:introduction}
\vspace{-.5\baselineskip}
Deep learning, which exploded from image processing, has now spread to audio processing.
However, in contrast to image processing, where the raw pixel data is commonly used,
different input representations are still used for different tasks in audio processing.
Log-amplitude mel spectrograms are widely used as one of the input features
in many tasks such as speech recognition, speech synthesis, and audio classification~\cite{Zhang2020-fe,Shen2018-xn,Heittola2020-gb}.
On the other hand, for extracting clean waveforms from a mixed waveform,
such as speech enhancement and source separation,
complex spectrograms (the outputs of STFT)
and raw waveforms are currently the mainstream~\cite{Luo2019-jx,Hu2020-rn}.

Phase information is present in the waveforms and complex spectrograms
but missing in the log-amplitude mel spectrograms.
Phase is more challenging to handle and interpret than the amplitude,
and it has been mentioned that the phase of the complex spectrum is not important in speech enhancement~\cite{Wang1982-up}.
However, as mentioned above, in speech enhancement, complex spectrograms
are increasingly being used as a more effective representation than amplitude spectrograms~\cite{Tan2019-gk,Pandey2019-oc,Hu2020-nk}.
The instantaneous frequency, which is the time derivative of the phase of the complex spectrogram,
was used for F0 estimation~\cite{Kawahara2011-nz}.
The group delay, which is the frequency derivative of the phase,
was used for formant analysis~\cite{Murthy2011-um}.

Previous studies examined
handcrafted constant-Q transform features and raw STFT phase for their specific tasks~\cite{Yang2021-re,Loweimi2021-ri}.
Our study aimed to investigate the effectiveness of the phase
obtained from the time-frequency representation of a learnable filterbank
for various classification tasks.
Specifically, we compared the phase, group delay, and instantaneous frequency as phase information.
We evaluated the performance for eight tasks, including voice, musical, and environmental sounds.
Our code is publicly available~\cite{Hidaka}.

\begin{figure*}[t]
  \centering
  \includegraphics[width=.82\linewidth]{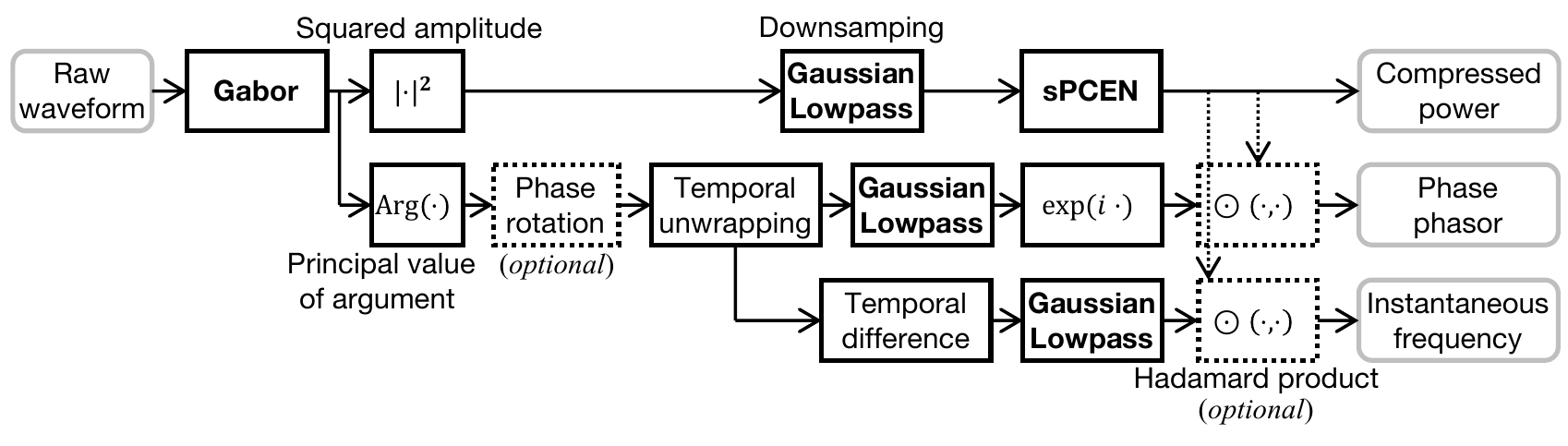}
  \vspace{-1.5\baselineskip}
  \caption{LEAF-extended.
  To eliminate redundancy, the group delay calculation path is omitted from the figure.
  The group delay is calculated by changing the temporal unwrapping and difference in the instantaneous frequency path
  to frequency unwrapping and difference, respectively.}
  \label{fig:leaf}
  \vspace{-1.3\baselineskip}
\end{figure*}

\section{METHODS}
\label{sec:methods}
\vspace{-.7\baselineskip}
When calculating a log-amplitude mel spectrogram,
a logarithmic conversion of the frequency axis is performed after discarding the phase of the complex spectrogram~\cite{McFee2020-as}.
Therefore, it is not obvious how the phase is logarithmically converted.
While log-amplitude mel spectrograms are still used,
there are also attempts to learn a more optimal time-frequency representation using DNNs.
It has been shown that
the performance of LEarnable Audio Frontend (LEAF) proposed in one of those studies is
comparable to or even better than the log-amplitude mel spectrogram~\cite{Zeghidour2021-sa}.
In addition, LEAF can directly compute the complex time-frequency representation
with a logarithmic frequency axis.
Therefore, we investigated a LEAF variant that can add phase information,
which we call LEAF-extended,
and used it instead of the log-amplitude mel spectrogram.
Note that the frequency axis of the LEAF-extended is not completely logarithmic after learning,
as described below.

\vspace{-.3\baselineskip}
\subsection{Original LEAF: calculation path of amplitude information}
\label{sec:original_leaf}
\vspace{-.3\baselineskip}
Fig.~\ref{fig:leaf} shows LEAF-extended.
In this section, we first describe only the components included in the original LEAF.
The original LEAF has no phase computation path and only computes the compressed power.
The original LEAF consisted of a cascade of the following components:
a Gabor filterbank responsible for computing the time-frequency representation from raw waveforms and also for converting the frequency axis;
a calculator of squared amplitude;
a Gaussian lowpass filterbank responsible for the downsampling along the time axis;
an sPCEN compressor, a modified version of Per-Channel Energy Normalization (PCEN)~\cite{Wang2017-fo}, responsible for the nonlinear compression of the amplitude.
Except for the calculator of squared amplitude,
each component has some learnable parameters for each frequency bin.
Henceforth, let $M \in \N$ denote the total number of frequency bins of the LEAF,
and $W \in \N$ denote the window width of the LEAF.

The Gabor filterbank $(\varphi_m(n))_{m=1,\dots,M}$ has two kinds of learnable parameters:
the center frequencies $(\eta_m)_{m=1,\dots,M}$
and the inverse window widths $(\sigma_m)_{m=1,\dots,M}$.
This filterbank is defined by
\begin{equation}
  \begin{gathered}
    \varphi_m(n) = \varphi_m'(n)/\Verts{\varphi'_m}_{l_1},\;
    \varphi_m'(n) = e^{\parens{-\frac{n^2}{2\sigma_m^2} + 2 \pi i \eta_m n}}, \\
    m = 1,\dots,M,\quad
    n = -W/2,\dots,W/2,
  \end{gathered}
\end{equation}
where $i = \sqrt{-1}$ is the imaginary unit, and $n, m \in \N$ are the time and frequency indices, respectively.\footnote{\ninept
  In the original paper~\cite{Zeghidour2021-sa},
  the $l_1$ normalization of each filter is done with $\varphi_m(n) = \varphi_m'(n)/\sqrt{2\pi}\sigma_m$.
  However, this normalization is not exact because the width of the Gabor filter is finite, $W+1$.
}
The learnable parameters are initialized
so that the frequency response has a similar shape as the mel filterbank.
The filterbank is regarded as a kernel where each filter is a channel,
and 1-D convolution of the filterbank and the raw signal yields the time-frequency representation.
The stride of the convolution is one.
If $(\eta_m)_{m=1,\dots,M}$ are linearly equally spaced
and $(\sigma_m)_{m=1,\dots,M}$ are constant,
the convolution corresponds to STFT using a Gaussian window.
Note that if the center frequencies $(\eta_m)_{m=1,\dots,M}$ become not monotonically increasing due to learning,
they are forced to reorder $(\eta_m)_{m=1,\dots,M}$ so that they become monotonically increasing.

The Gaussian lowpass filterbank $(\phi_m(n))_{m=1,\dots,M}$
has inverse window width parameters $(\sigma_m)_{m=1,\dots,M}$ as learnable parameters.
This filterbank is defined by
\begin{equation}
  \begin{gathered}
    \phi_m(n) = \phi_m'(n)/\Verts{\phi'_m}_{l_1},\quad
    \phi_m'(n) = e^{-\frac{n^2}{2(0.5\sigma_m(W-1))^2}}, \\
    m = 1,\dots,M,\quad
    n = -W/2,\dots,W/2.
  \end{gathered}
\end{equation}
Depthwise 1D convolution, whose stride is greater than one,
of the Gaussian lowpass filterbank and the squared amplitude of the Gabor filterbank output
yields a downsampled representation.\footnote{\ninept
  In the original paper~\cite{Zeghidour2021-sa},
  the $l_1$ normalization of each filter of the Gaussian lowpass filterbank is performed like the Gabor filterbank.
}

The sPCEN compressor $\operatorname{PCEN}(\mathcal{F}(m,n))$
is a learnable DNN component derived from
PCEN, a method for nonlinear compression of amplitude per channel~\cite{Wang2017-fo}.
This component is defined by
\begin{equation}
  \begin{gathered}
    \operatorname{PCEN}(\mathcal{F}(m,n)) = \parens{\frac{\mathcal{F}(m,n)}{(\epsilon + \mathcal{M}(m,n))^{a_m}} + \delta_m}^{1/r_m} - \delta_m^{1/r_m}, \\
    \mathcal{M}(m,n) = (1-s_m)\mathcal{M}(m,n-1)+s_m\mathcal{F}(m,n),\\
    m = 1,\dots,M,\quad
    n = 1,\dots,L,
  \end{gathered}
\end{equation}
where $\mathcal{F}(m,n)$ is the time-frequency representation that is the output of the Gaussian lowpass filterbank;
$L$ is the temporal length of $\mathcal{F}(m,n)$;
$\epsilon$ is a small value that avoids zero division;
and $(\alpha_m)_{m=1,\dots,M}$, $(\delta_m)_{m=1,\dots,M}$, $(r_m)_{m=1,\dots,M}$, $(s_m)_{m=1,\dots,M}$ are the learnable parameters of the sPCEN compressor.
Henceforth, the power compressed by sPCEN is denoted as $\mathbf{POW}$.

\vspace{-.3\baselineskip}
\subsection{Two definitions of phase}
\label{sec:definitions_phase}
\vspace{-.3\baselineskip}
We will now pause the description of LEAF-extended
to touch on two definitions of continuous STFT with different phases~\cite{Yatabe2019-vb}.
STFT of a time-domain signal $x(t)$ for $w(t) \neq 0$ can be expressed by
\begin{equation}
  \operatorname{STFT_1}(f,t)
  = \int_\R x(\tau)\overline{w(\tau-t)e^{2\pi if(\tau-t)}}d\tau,
\end{equation}
where $\overline{z}$ is the complex conjugate of $z$,
and $t, f \in \R$ are the time and frequency, respectively.
The other definition is expressed by
\begin{equation}
  \operatorname{STFT_2}(f,t)
  = \int_\R x(\tau)\overline{w(\tau-t)e^{2\pi if\tau}}d\tau.
\end{equation}
In the first definition,
the term of the complex sinusoid moves with the window.
On the other hand, in the second definition,
the term is fixed to the origin.
The difference between these two definitions yields the following relationship between their phase values:
\begin{equation}
  \angle\operatorname{STFT_1}(f,t)
  = \angle\operatorname{STFT_2}(f,t) + 2\pi ft.
\end{equation}
This yields the following relationship between the instantaneous frequencies,
which are the time derivatives of the phases,
and the group delays,
which are the negative frequency derivatives of the phases:
\begin{align}
  (\partial/\partial t)\angle\operatorname{STFT_1}(f,t)
  &= (\partial/\partial t)\angle\operatorname{STFT_2}(f,t) + 2\pi f, \\
  -(\partial/\partial f)\angle\operatorname{STFT_1}(f,t)
  &= -(\partial/\partial f)\angle\operatorname{STFT_2}(f,t) - 2\pi t. \label{eq:gd}
\end{align}
$(\partial/\partial t)\angle\operatorname{STFT_1}(f,t)$ is the more accepted definition of ``instantaneous frequency''
because it represents the absolute frequency.
On the other hand, $(\partial/\partial t)\angle\operatorname{STFT_2}(f,t)$ is often called ``relative instantaneous frequency''
because it represents the frequency relative to the (angular) frequency axis $2\pi f$.
$-(\partial/\partial f)\angle\operatorname{STFT_1}(f,t)$ is referred to as the ``group delay.''
Since $-(\partial/\partial f)\angle\operatorname{STFT_2}(f,t)$ varies linearly with the time position,
to our knowledge, it has not been actively used in signal processing.
We enabled LEAF-extended to handle the different phase definitions
and examined their effects.

\begin{figure*}[t]
  \centering
  \includegraphics[width=.91\linewidth]{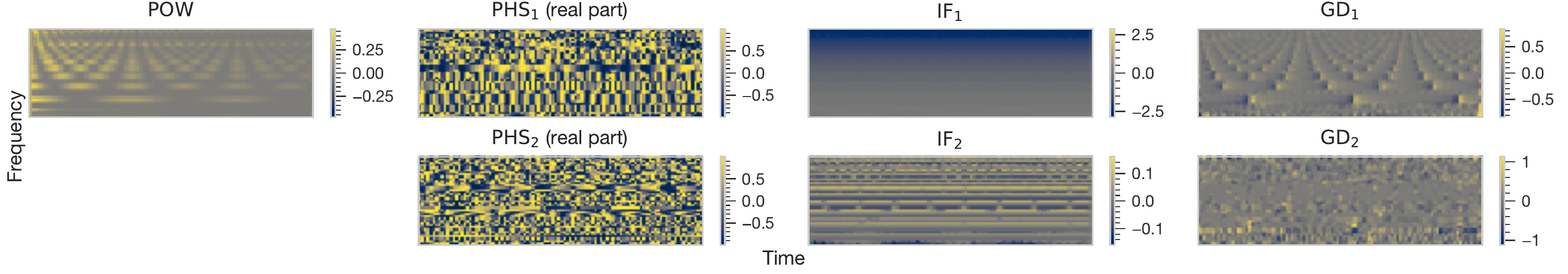}
  \vspace{-1.4\baselineskip}
  \caption{The heatmaps of the output of the untrained LEAF-extended.
  Here, a musical note of an electronic keyboard with a pitch of C5
  (``keyboard\_electronic\_012-072-127.wav'' from the NSynth training set)
  was used as an input.}
  \label{fig:features}
  \vspace{-1.65\baselineskip}
\end{figure*}

\vspace{-.3\baselineskip}
\subsection{LEAF-extended: calculation paths of phase information}
\label{sec:leaf_extended}
\vspace{-.3\baselineskip}
This section describes the calculation paths for phase information in LEAF-extended.
As already mentioned, recall that Fig.~\ref{fig:leaf} shows LEAF-extended.
Note that the terms phase, instantaneous frequency, and group delay
are abused in this section without paying attention to their signs.
Even if the sign is reversed from the original definition of STFT,
it is only necessary to reverse the sign of the weights of the first linear layer that receives the LEAF-extended output.
Therefore, signs do not affect the training of the neural network.

First, let us describe the calculation path for the phasor of the phase.
In this path, the principal value of the argument $\theta_1 \in (-\pi, \pi]$ is firstly calculated from the complex time-frequency representation,
which is the output of the Gabor filterbank.
This argument $\theta_1$ corresponds to the phase in the sense of $\angle\operatorname{STFT_1}$.
Optionally, this argument $\theta_1$ is converted to another argument $\theta_2 \in (-\pi, \pi]$ in the sense of $\angle\operatorname{STFT_2}$ by rotating $\theta_1$ as follows:
\begin{gather}
  \theta_2(m,n) = p(\theta_1(m,n) + \eta_m n),
\end{gather}
where $(\eta_m)_{m=1,\dots,M}$ are the center frequencies of the Gabor filterbank,
$p(\theta) := \theta_\mathrm{principal}$ maps an angular $\theta \in \R$ to its principal value $\theta_\mathrm{principal} \in (-\pi, \pi]$.
Then, the phase is unwrapped in the time direction
and downsampled by a Gaussian lowpass filterbank for the phase.\footnote{\ninept
  If the $l_1$-norm of the Gaussian lowpass filterbank is not strictly normalized to 1,
  the downsampling causes the phase scale to change.
  The scale change results in an undesired phase shift.
}
Finally, the downsampled phase is converted to a phasor
by the function $f(\theta) := e^{i\theta}$.
Note that the phasor is regarded as a two-channel image with real and imaginary parts.
Henceforth, the phasors via the path without/with the phase rotation are denoted as $\mathbf{PHS}_1$ and $\mathbf{PHS}_2$, respectively.

Next, we will explain the path for calculating the instantaneous frequency.
The first step in this path is to calculate the argument $\theta_1$ or its rotated version $\theta_2$.
Then, after unwrapping it in the time direction,
the instantaneous frequency is calculated by taking the difference in the time direction.
Finally, it is downsampled by a Gaussian lowpass filterbank for the instantaneous frequency.
Henceforth, the instantaneous frequencies via the path without/with the phase rotation are denoted as $\mathbf{IF}_1$ and $\mathbf{IF}_2$, respectively.

The group delay is calculated by changing the temporal unwrapping and difference in the instantaneous frequency path
to frequency unwrapping and difference, respectively.
Henceforth, the group delays via the path without/with the phase rotation are denoted as $\mathbf{GD}_1$ and $\mathbf{GD}_2$, respectively.

Optionally, the phase features are elementwise multiplied by $\mathbf{POW}$.
This elementwise multiplication is expected to weaken the influence of the phase features of elements with low power.

Fig.~\ref{fig:features} shows the features output by LEAF-extended.
In the frequency bands where a sinusoidal component exists,
the values of $\textbf{PHS}_1$ align and rotate,
whereas this is not the case for $\textbf{PHS}_2$.
$\mathbf{IF}_1$ has a bias along the frequency axis,
while $\mathbf{IF}_2$ has an approximately uniform distribution of values
over the entire frequency range.
$\mathbf{GD}_1$ shows a characteristic pattern at the inflection points of the amplitude,
while $\mathbf{GD}_2$ is conspicuous by its otherwise noisier pattern.
According to Eq.~(\ref{eq:gd}), some might expect $\mathbf{GD}_2$ to be represented as $\mathbf{GD}_1$ with a linear bias in the time direction.
However, $\mathbf{GD}_2$ presents a messy pattern
because $\mathbf{GD}_1$ and $\mathbf{GD}_2$ are approximations by taking the difference rather than analytical derivatives
and because the phase cycle with a period of $2\pi$.

\vspace{-.7\baselineskip}
\subsection{Treatment of the phase of zero amplitude elements}
\vspace{-.3\baselineskip}
The phase is not defined for elements with zero amplitude.
In our implementation, the instantaneous frequencies and group delays calculated for undefined phase elements
are also regarded as undefined.
Undefined values are replaced by linear interpolation during downsampling with a Gaussian lowpass filterbank.
If the value of the central element of the sampling was undefined,
the value of the corresponding element after sampling is also considered undefined.
Finally, undefined values are replaced with $0$ ($0+0i$ for $\mathbf{PHS}_1$ and $\mathbf{PHS}_2$).
In addition, the backpropagation of the argument of the elements with zero amplitude is set to zero.

\begin{table*}[t]
    \centering
    \caption{
      Test accuracy (\%) for audio classification.
      The 95\% confidence intervals were calculated from the sample sizes of the datasets.
      \textbf{Bold scores} are significantly higher than the $\mathbf{POW}$ scores.
      \underline{Underlined scores} are significantly lower than the $\mathbf{POW}$ scores.
    }
    \label{tbl:results}
    \footnotesize
    \begin{tabular}{lcccccccc}
      \toprule
      Feature                               &            Music (pitch) &               Music (inst.) &                    Language &                     Speaker &                    Birdsong &          Scenes &                     Keyword &         Emotion \\
      \midrule
      $\mathbf{POW}$                        &           $91.5 \pm 0.4$ &              $69.2 \pm 0.7$ &              $74.7 \pm 0.5$ &              $35.5 \pm 0.6$ &              $76.4 \pm 0.9$ &  $66.2 \pm 1.8$ &              $94.1 \pm 0.3$ &  $60.2 \pm 2.0$ \\
      \midrule
      $+ \mathbf{PHS}_1$                    &  $\mathbf{92.9 \pm 0.4}$ &     $\mathbf{71.5 \pm 0.7}$ &     $\mathbf{79.3 \pm 0.5}$ &              $36.1 \pm 0.6$ &              $75.1 \pm 0.9$ &  $66.7 \pm 1.8$ &              $94.4 \pm 0.3$ &  $60.7 \pm 2.0$ \\
      $+ \mathbf{PHS}_1 \odot \mathbf{POW}$ &           $91.9 \pm 0.4$ &              $69.6 \pm 0.7$ &              $74.5 \pm 0.5$ &              $35.6 \pm 0.6$ &              $75.6 \pm 0.9$ &  $68.8 \pm 1.8$ &              $94.1 \pm 0.3$ &  $62.0 \pm 2.0$ \\
      $+ \mathbf{PHS}_2$                    &  $\mathbf{92.7 \pm 0.4}$ &              $68.0 \pm 0.7$ &  $\underline{72.8 \pm 0.5}$ &     $\mathbf{37.4 \pm 0.6}$ &              $74.9 \pm 1.0$ &  $65.3 \pm 1.9$ &              $94.4 \pm 0.3$ &  $60.8 \pm 2.0$ \\
      $+ \mathbf{PHS}_2 \odot \mathbf{POW}$ &           $91.9 \pm 0.4$ &              $69.0 \pm 0.7$ &              $74.0 \pm 0.5$ &     $\mathbf{37.1 \pm 0.6}$ &              $74.8 \pm 1.0$ &  $65.5 \pm 1.9$ &              $94.2 \pm 0.3$ &  $61.1 \pm 2.0$ \\
      \midrule
      $+ \mathbf{IF}_1$                     &  $\mathbf{93.2 \pm 0.4}$ &              $69.9 \pm 0.7$ &  $\underline{71.3 \pm 0.5}$ &              $34.6 \pm 0.6$ &              $76.0 \pm 0.9$ &  $69.4 \pm 1.8$ &              $94.2 \pm 0.3$ &  $57.3 \pm 2.0$ \\
      $+ \mathbf{IF}_1 \odot \mathbf{POW}$  &           $92.1 \pm 0.4$ &              $69.5 \pm 0.7$ &     $\mathbf{78.1 \pm 0.5}$ &              $35.0 \pm 0.6$ &              $76.2 \pm 0.9$ &  $65.9 \pm 1.9$ &              $94.0 \pm 0.3$ &  $61.0 \pm 2.0$ \\
      $+ \mathbf{IF}_2$                     &  $\mathbf{93.2 \pm 0.4}$ &              $70.2 \pm 0.7$ &  $\underline{49.2 \pm 0.6}$ &     $\mathbf{37.2 \pm 0.6}$ &  $\underline{73.7 \pm 1.0}$ &  $68.2 \pm 1.8$ &              $94.5 \pm 0.3$ &  $57.5 \pm 2.0$ \\
      $+ \mathbf{IF}_2 \odot \mathbf{POW}$  &  $\mathbf{93.0 \pm 0.4}$ &              $69.9 \pm 0.7$ &  $\underline{68.3 \pm 0.5}$ &              $36.1 \pm 0.6$ &  $\underline{73.5 \pm 1.0}$ &  $66.8 \pm 1.8$ &              $94.4 \pm 0.3$ &  $60.9 \pm 2.0$ \\
      \midrule
      $+ \mathbf{GD}_1$                     &           $91.3 \pm 0.4$ &     $\mathbf{72.6 \pm 0.7}$ &     $\mathbf{83.9 \pm 0.4}$ &              $35.4 \pm 0.6$ &     $\mathbf{79.8 \pm 0.9}$ &  $68.2 \pm 1.8$ &              $94.4 \pm 0.3$ &  $58.7 \pm 2.0$ \\
      $+ \mathbf{GD}_1 \odot \mathbf{POW}$  &  $\mathbf{92.5 \pm 0.4}$ &  $\underline{67.7 \pm 0.7}$ &     $\mathbf{79.4 \pm 0.5}$ &     $\mathbf{37.6 \pm 0.6}$ &              $77.0 \pm 0.9$ &  $67.2 \pm 1.8$ &              $94.4 \pm 0.3$ &  $60.5 \pm 2.0$ \\
      $+ \mathbf{GD}_2$                     &  $\mathbf{92.8 \pm 0.4}$ &              $69.2 \pm 0.7$ &  $\underline{70.5 \pm 0.5}$ &  $\underline{32.8 \pm 0.6}$ &              $77.4 \pm 0.9$ &  $66.4 \pm 1.8$ &  $\underline{92.8 \pm 0.3}$ &  $60.6 \pm 2.0$ \\
      $+ \mathbf{GD}_2 \odot \mathbf{POW}$  &           $92.3 \pm 0.4$ &              $68.3 \pm 0.7$ &     $\mathbf{76.1 \pm 0.5}$ &  $\underline{32.0 \pm 0.6}$ &              $77.5 \pm 0.9$ &  $68.5 \pm 1.8$ &              $93.6 \pm 0.3$ &  $57.9 \pm 2.0$ \\
      \bottomrule
    \end{tabular}
    \vspace{-1.9\baselineskip}
\end{table*}

\vspace{-.5\baselineskip}
\section{EXPERIMENTAL RESULTS}
\vspace{-.5\baselineskip}

\subsection{Experiments}
\vspace{-.3\baselineskip}
In this study, we investigated whether the addition of phase features contributes to performance improvement for eight audio classification tasks:
musical pitch and instrument detection on NSynth~\cite{Engel2017-uq} (112 pitches, 11 instruments, 289,205 training samples, 16,774 evaluation samples),
language identification on VoxForge~\cite{Revay2019-qs} (6 classes, 148,654 training samples, 27,764 evaluation samples),
speaker identification on VoxCeleb~\cite{Nagrani2017-tf} (1,251 classes, 128,086 training samples, 25,430 evaluation samples),
birdsong detection on DCASE2018~\cite{Stowell2018-yj} (2 classes, 35,690 training samples, 12,620 evaluation samples),
acoustic scene classification on TUT~\cite{Heittola2018-mk} (10 classes, 6,122 training samples, 2,518 evaluation samples),
keyword spotting on SpeechCommands~\cite{Warden2018-wa} (35 classes, 84,843 training samples, 20,986 evaluation samples),
emotion recognition on CREMA-D~\cite{Cao2014-vx} (6 classes, 5,146 training samples, 2,296 evaluation samples).
These datasets are the same as those used in the original LEAF paper~\cite{Zeghidour2021-sa}.
The sampling frequency was set to 16 kHz.
We ensured that
the recording conditions were not shared between the training and evaluation sets
for datasets with various recording conditions.
See our implementation for details on splitting the datasets~\cite{Hidaka}.

As the DNN structure for these audio classification tasks,
we used a cascade of LEAF-extended,
2-D Batch Normalization~\cite{Ioffe2015-hn},
and EfficientNetB0 \cite{Tan2019-ip},
a CNN with about 4 million parameters.
The power and phase features of LEAF-extended were stacked in the channel direction
like a 2-D image.
For $\mathbf{PHS}_1$ and $\mathbf{PHS}_2$, after applying complex Batch Normalization~\cite{Trabelsi2018-ye},
the real and imaginary parts were decomposed and regarded as a real image consisting of two channels.
The window width $W$ and the number of frequency bins $M$ of a LEAF-extended were set to 40 and 401.
The $\sigma_m$ of a Gaussian lowpass filterbank was initialized to 0.4, and its stride was set to 100.
The $\alpha_m,\delta_m,r_m,s_m$ of an sPCEN compressor were initialized to 0.96, 2.0, 2.0, 0.04 respectively,
and the $\epsilon$ to prevent zero division was set to $5\times10^{-8}$.

During the training, each sample was randomly cut into a one-second interval and used as an input.
During the evaluation, each sample was cut into one-second intervals without overlap,
and the average of all logits from the DNN was used for the final prediction output.
The batch size was set to 256.
Learning was terminated when the validation loss did not improve
after 50,000 consecutive steps for birdsong detection, language identification, and musical instrument detection
and after 20,000 consecutive steps for the other tasks.
We trained the networks by using Adam~\cite{Kingma2014-rm} as an optimizer.
To suppress overfitting, the following methods were used:
SpechAugment~\cite{Park2019-ik} (W: 0, F: 5, T: 11, times of temporal and frequency masking: 2 times each) for the input to EfficientNetB0,
$l_2$ regularization (weight: 0.00001) for the convolutional layer of the EfficientNetB0,
Dropout~\cite{Srivastava2014-cl} (rate: 0.2) for the last fully-connected layer of the EfficientNetB0,
and label smoothing~\cite{Muller2019-xj} (rate: 0.1) for the true labels.

\begin{figure}[t]
  \centering
  \includegraphics[width=\linewidth]{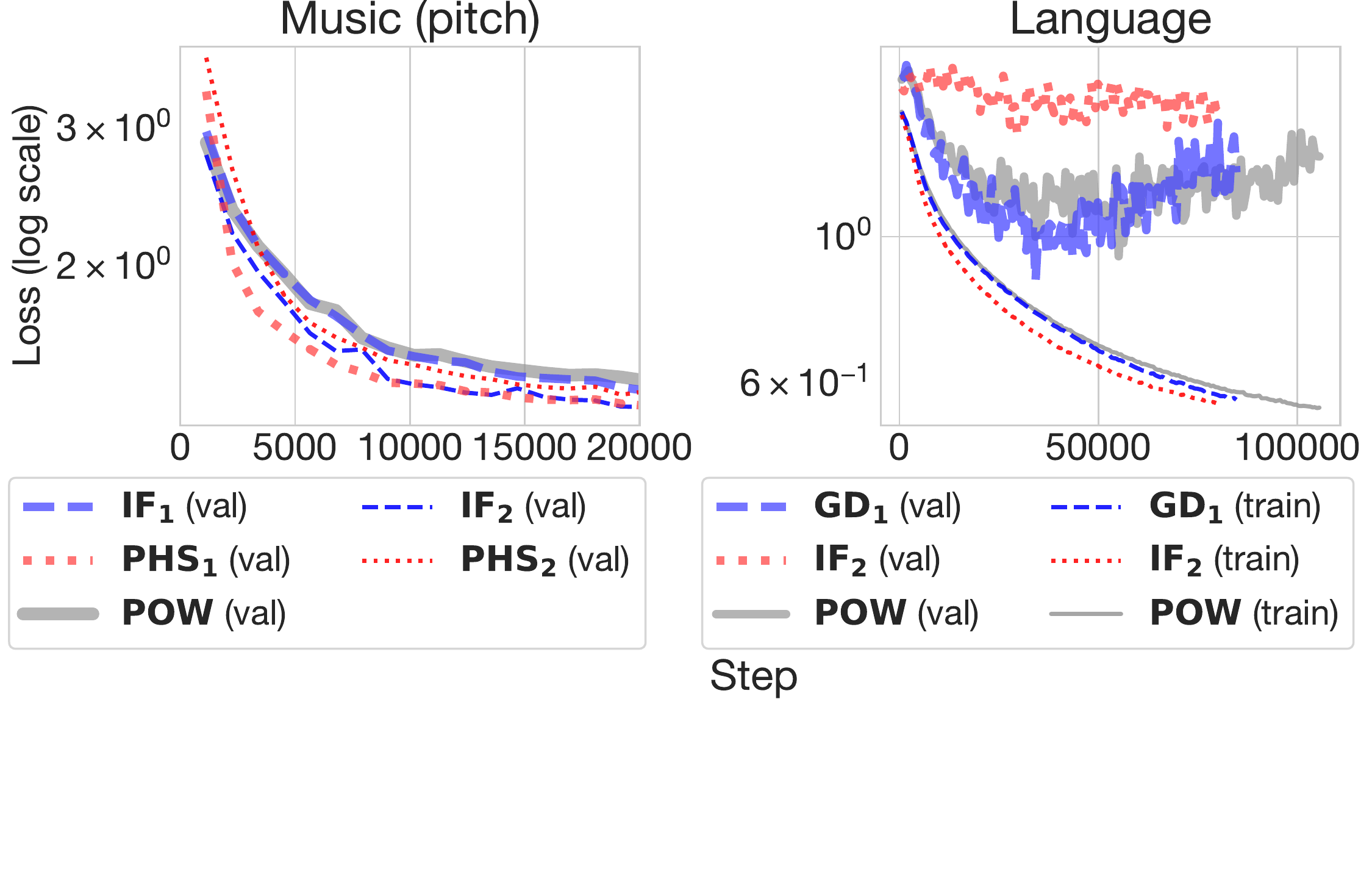}
  \vspace{-2.1\baselineskip}
  \caption{Learning curves}
  \label{fig:learning_curves}
  \vspace{-1.5\baselineskip}
\end{figure}

\vspace{-.8\baselineskip}
\subsection{Results and discussion}
\vspace{-.4\baselineskip}
Table~\ref{tbl:results} shows the experimental results.
For each task and each feature, DNN was independently trained.
Before dealing with the individual tasks, we will first discuss some of the overall tendencies.
For a given task, if the phase phasor significantly improved performance,
then the derivatives of the phase always significantly improved performance as well.
This fact suggests that in audio classification,
the relationship between adjacent elements of the phase is more important
than the phase value itself.
The elementwise multiplication with $\mathbf{POW}$ sometimes prevented either significant performance degradation or improvement.
While the multiplication suppresses unwanted information, the physical meaning of the phase features might be corrupted.
This corruption might be solved by convolution on the phase features without the elementwise multiplication
and then applying the gating mechanism~\cite{Dauphin2017-jq}.
$\mathbf{GD}_1$ tended to be superior to $\mathbf{GD}_2$,
and $\mathbf{GD}_2$ often resulted in significant performance degradation.
As mentioned in Section~\ref{sec:leaf_extended},
we think the degradation was caused by the fact that $\mathbf{GD}_2$ produces noisy and non-essential patterns in audio data.
The inferiority of $\mathbf{GD}_2$ also implies that it is difficult to obtain valuable information about the group delay from $\mathbf{PHS}_2$.

For the musical pitch detection, most of the phase features, especially the instantaneous frequency, performed well.
This result corresponds to the fact that
the instantaneous frequency has already been applied to F0 estimation successfully~\cite{Kawahara2011-nz}.
Fig.~\ref{fig:learning_curves} (left) shows the learning curves of some features.
$\mathbf{IF}_2$ converged faster than $\mathbf{IF}_1$.
On the frequency axis, the distribution of $\mathbf{IF}_1$ is skewed, while that of $\mathbf{IF}_2$ is not.
$\mathbf{IF}_2$ could be a more appropriate feature for CNNs that capture local features,
as the power normalization of each frequency bin leads to better performance~\cite{Nagrani2017-tf}.
In contrast, $\mathbf{PHS}_1$ converged faster than $\mathbf{PHS}_2$.
Probably, this is because
the property of $\textbf{PHS}_1$ that the values align and rotate in the frequency bands where a sinusoidal component exists, as shown in Fig.~\ref{fig:features}, was easily captured by the CNN.

For the musical instrument detection, $\mathbf{GD}_1$ and $\mathbf{PHS}_1$ showed significant improvement.
As shown in Fig.~\ref{fig:features}, the group delay is considered to reflect the timbre characteristics to some extent.

For the language identification, some phase features, especially $\mathbf{GD}_1$, performed well.
In contrast, some other phase features, especially $\mathbf{IF}_2$, resulted in significant performance degradation.
Fig.~\ref{fig:learning_curves} (right) shows the learning curves of $\mathbf{GD}_1$, $\mathbf{IF}_2$, and $\mathbf{POW}$.
For $\mathbf{GD}_1$, both training loss and validation loss decreased faster than $\mathbf{POW}$.
On the other hand, $\mathbf{IF}_2$ showed the fastest decrease in training loss among all the features
but almost no decrease in validation loss.
Voxforge consists of speech data from many speakers, and their recording conditions are various.
In our experiments, we split the training and the validation set so that the speakers are different.
Therefore, it is possible that $\mathbf{IF}_2$ actively acquired information about the recording conditions rather than linguistic information.

For the speaker identification, $\mathbf{PHS}_2$, $\mathbf{IF}_2$, and $\mathbf{GD}_1$ performed significantly.
The group delay has been applied to formant estimation~\cite{Murthy2011-um}, 
and $\mathbf{GD}_1$ is considered to have acquired information about individuality.
The recording conditions (video IDs) were not overlapped between training and validation sets.
Nevertheless, it is possible that $\mathbf{PHS}_2$ and $\mathbf{IF}_2$ gained some valuable information.

For the birdsong detection, $\mathbf{IF}_2$ showed significant performance degradation.
DCASE2018 dataset is a collection of three independent datasets,
two of which were used as the training set and the rest as the validation set.
For example, many audio samples in BirdVox, one of the datasets comprising the training set, contain power line hum.
Therefore, as in the language identification,
probably, this degradation was caused by $\mathbf{IF}_2$ promoting overfitting of the recording conditions.

For the acoustic scene classification, keyword spotting, and emotion recognition,
the phase features did not significantly affect in most cases.
The phase features might not be necessary for these tasks.
Alternatively, the sample sizes of the datasets might be too small to show significant differences.

Overall, $\mathbf{PHS}_1$ as phase phasor, $\mathbf{IF}_2$ as instantaneous frequency, and $\mathbf{GD}_1$ as group delay
tended to be efficient for further feature acquisition.
While $\mathbf{PHS}_1$ and $\mathbf{GD}_1$ resulted in generally better performance,
$\mathbf{IF}_2$ resulted in both better and worse performance.
Presumably, the performance degradation happened because $\mathbf{IF}_2$ led to overfitting of the recording conditions of the training sets.
There is a possibility that
domain-adversarial training~\cite{Ganin2016-ao} could prevent the overfitting of recording conditions.

\vspace{-.8\baselineskip}
\section{CONCLUSIONS}
\vspace{-.7\baselineskip}
In this paper, we investigated the effectiveness of phase features of a time-frequency representation for audio classification.
The results suggested that the phase and its derivatives are valuable in some classification tasks.
Future work should address the impact of recording conditions
and exploit gating and domain-adversarial mechanisms.

\textbf{Acknowledgment:} This work was supported by JST SPRING, Grant Number JPMJSP2136.



\begin{thebibliography}{99}
\bibitem{Zhang2020-fe}
Y.~Zhang, J.~Qin, D.~S.~Park, W.~Han, C.~C.~Chiu, R.~Pang,
Q.~V.~Le, and Y.~Wu,
\newblock ``Pushing the limits of {Semi-Supervised} learning for automatic
speech recognition,''
\newblock {\em arXiv preprint arXiv:2010.10504}, 2020.

\bibitem{Shen2018-xn}
J.~Shen, R.~Pang, R.~J.~Weiss, M.~Schuster, N.~Jaitly,
Z.~Yang, Z.~Chen, Y.~Zhang, Y.~Wang, R.~Ryan, R.~A.~Saurous,
Y.~Agiomvrgiannakis, and Y.~Wu,
\newblock ``Natural {TTS} synthesis by conditioning wavenet on {MEL}
spectrogram predictions,''
\newblock in {\em ICASSP}, 2018, pp. 4779--4783.

\bibitem{Heittola2020-gb}
T.~Heittola, A.~Mesaros, and T.~Virtanen,
\newblock ``Acoustic scene classification in {DCASE} 2020 challenge:
generalization across devices and low complexity solutions,''
\newblock {\em arXiv preprint arXiv:2005.14623}, 2020.

\bibitem{Luo2019-jx}
Y.~Luo and N.~Mesgarani,
\newblock ``{Conv-TasNet}: Surpassing ideal {Time--Frequency} magnitude masking
for speech separation,''
\newblock {\em IEEE/ACM Trans. on Audio, Speech, and Language Processing},
vol. 27, no. 8, pp. 1256--1266, 2019.

\bibitem{Hu2020-rn}
S.~Hu, B.~Zhang, B.~Liang, E.~Zhao, S.~Lui, T.~Music, and E.~Tme,
\newblock ``Phase-aware music super-resolution using generative adversarial
networks,''
\newblock in {\em {INTERSPEECH}}, 2020, pp. 4074--4078.

\bibitem{Wang1982-up}
D.~Wang and J.~Lim,
\newblock ``The unimportance of phase in speech enhancement,''
\newblock {\em IEEE Trans. Acoust.}, vol. 30, no. 4, pp. 679--681, 1982.

\bibitem{Tan2019-gk}
K.~Tan and D.~Wang,
\newblock ``Complex Spectral Mapping with a Convolutional Recurrent Network for Monaural Speech Enhancement,''
\newblock in {\em ICASSP}, 2019, pp. 6865–6869.

\bibitem{Pandey2019-oc}
A.~Pandey and D.~Wang,
\newblock ``Exploring deep complex networks for complex spectrogram
enhancement,''
\newblock in {\em ICASSP}, 2019, pp. 6885--6889.

\bibitem{Hu2020-nk}
Y.~Hu, Y.~Liu, S.~Lv, M.~Xing, S.~Zhang, Y.~Fu, J.~Wu,
B.~Zhang, and L.~Xie,
\newblock ``{DCCRN} : Deep complex convolution recurrent network for
{Phase-Aware} speech enhancement,''
\newblock in {\em {INTERSPEECH}}, 2020, pp. 2472--2476.

\bibitem{Kawahara2011-nz}
H.~Kawahara, T.~Irino, and M.~Morise,
\newblock ``An interference-free representation of instantaneous frequency of
periodic signals and its application to {F0} extraction,''
\newblock in {\em ICASSP}, 2011, pp. 5420--5423.

\bibitem{Murthy2011-um}
H.~A.~Murthy and B.~Yegnanarayana,
\newblock ``Group delay functions and its applications in speech technology,''
\newblock {\em Sadhana - Academy Proceedings in Engineering Sciences}, vol. 36,
no. 5, pp. 745--782, 2011.

\bibitem{Yang2021-re}
J.~Yang, H.~Wang, R.~K.~Das, and Y.~Qian,
\newblock ``Modified Magnitude-Phase Spectrum Information for Spoofing Detection,''
\newblock {\em IEEE/ACM Trans. on Audio, Speech, and Language Processing},
vol. 29, pp. 1065--1078, 2021.

\bibitem{Loweimi2021-ri}
E.~Loweimi, Z.~Cvetkovic, P.~Bell, and S.~Renals,
\newblock ``Speech Acoustic Modelling from Raw Phase Spectrum,''
\newblock in {\em ICASSP}, 2021, pp. 6738--6742.

\bibitem{Hidaka}
https://github.com/onkyo14taro/investigation-phase.

\bibitem{McFee2020-as}
B.~McFee {\em et al},
\newblock ``librosa/librosa: 0.8.0,'' https://doi.org/10.5281/zenodo.3955228, 2020.

\bibitem{Zeghidour2021-sa}
N.~Zeghidour, O.~Teboul, F.~C.~Quitry, and M.~Tagliasacchi,
\newblock ``{LEAF}: A learnable frontend for audio classification,''
\newblock in {\em ICLR}, 2021.

\bibitem{Wang2017-fo}
Y.~Wang, P.~Getreuer, T.~Hughes, R.~F.~Lyon, and R.~A.~Saurous,
\newblock ``Trainable frontend for robust and far-field keyword spotting,''
\newblock in {\em ICASSP}, 2017, pp. 5670--5674.

\bibitem{Yatabe2019-vb}
K.~Yatabe, Y.~Masuyama, T.~Kusano, and Y.~Oikawa,
\newblock ``Representation of complex spectrogram via phase conversion,''
\newblock {\em Acoust. Sci. Technol.}, vol. 40, no. 3, pp. 170--177, 2019.

\bibitem{Engel2017-uq}
J.~Engel, C.~Resnick, A.~Roberts, S.~Dieleman, M.~Norouzi,
D.~Eck, and K.~Simonyan,
\newblock ``Neural audio synthesis of musical notes with wavenet
autoencoders,''
\newblock in {\em ICML}, 2017, pp. 1068--1077.

\bibitem{Revay2019-qs}
S.~Revay and M.~Teschke,
\newblock ``Multiclass language identification using deep learning on spectral
images of audio signals,''
\newblock {\em arXiv preprint arXiv:1905.04348}, 2019.

\bibitem{Nagrani2017-tf}
A.~Nagrani, J.~S.~Chung, and A.~Zisserman,
\newblock ``Voxceleb: a large-scale speaker identification dataset,''
\newblock {\em arXiv preprint arXiv:1706.08612}, 2017.

\bibitem{Stowell2018-yj}
D.~Stowell, M.~D.~Wood, H.~Pamu{\l}a, Y.~Stylianou, and H.~Glotin,
\newblock ``Automatic acoustic detection of birds through deep learning: The
first bird audio detection challenge,''
\newblock {\em Methods Ecol. Evol.}, vol. 10, no. 3, pp. 368--380, 2018.

\bibitem{Heittola2018-mk}
T.~Heittola, A.~Mesaros, and T.~Virtanen,
\newblock ``{TUT} urban acoustic scenes 2018, development dataset,'' 
https://doi.org/10.5281/zenodo.1228142, 2018.

\bibitem{Warden2018-wa}
P.~Warden,
\newblock ``Speech commands: A dataset for {Limited-Vocabulary} speech
recognition,''
\newblock {\em arXiv preprint arXiv:1804.03209}, 2018.

\bibitem{Cao2014-vx}
H.~Cao, D.~G.~Cooper, M.~K.~Keutmann, R.~C.~Gur, A.~Nenkova, and R.~Verma,
\newblock ``{CREMA-D}: Crowd-sourced emotional multimodal actors dataset,''
\newblock {\em IEEE Trans. Affect. Comput.}, vol. 5, no. 4, pp. 377--390, 2014.

\bibitem{Ioffe2015-hn}
S.~Ioffe and C.~Szegedy,
\newblock ``Batch normalization: Accelerating deep network training by reducing
internal covariate shift,''
\newblock in {\em ICML}, 2015, pp. 448--456.

\bibitem{Tan2019-ip}
M.~Tan and Q.~Le,
\newblock ``Efficientnet: Rethinking model scaling for convolutional neural
networks,''
\newblock in {\em ICML}, 2019, pp. 6105--6114.

\bibitem{Trabelsi2018-ye}
C.~Trabelsi, O.~Bilaniuk, D.~Serdyuk, S.~Subramanian,
J.~F.~Santos, S.~Mehri, N.~Rostamzadeh, Y.~Bengio, and C.~J.~Pal,
\newblock ``Deep complex networks,''
\newblock in {\em ICLR}, 2018.

\bibitem{Kingma2014-rm}
D.~P.~Kingma and J.~Ba,
\newblock ``Adam: A method for stochastic optimization,''
\newblock {\em arXiv preprint arXiv:1412.6980}, 2014.

\bibitem{Park2019-ik}
D.~S.~Park, W.~Chan, Y.~Zhang, C.~C.~Chiu, B.~Zoph, E.~D.~Cubuk, and Q.~V.~Le,
\newblock ``{SpecAugment}: A simple data augmentation method for automatic
speech recognition,''
\newblock in {\em INTERSPEECH}, 2019, pp. 2613--2617.

\bibitem{Srivastava2014-cl}
N.~Srivastava, G.~Hinton, A.~Krizhevsky, I.~Sutskever, and R.~Salakhutdinov,
\newblock ``Dropout: A simple way to prevent neural networks from
overfitting,''
\newblock {\em J. Mach. Learn. Res.}, vol. 15, no. 1, pp. 1929--1958, 2014.

\bibitem{Muller2019-xj}
R.~M{\"u}ller, S.~Kornblith, and G.~Hinton,
\newblock ``When does label smoothing help?,''
\newblock in {\em NeurIPS}, 2019.

\bibitem{Dauphin2017-jq}
Y.~N.~Dauphin, A.~Fan, M.~Auli, and D.~Grangier,
\newblock ``Language modeling with gated convolutional networks,''
\newblock in {\em ICML}, 2017, pp. 933--941.

\bibitem{Ganin2016-ao}
Y.~Ganin, E.~Ustinova, H.~Ajakan, P.~Germain, H.~Larochelle,
F.~Laviolette, M.~Marchand, and V.~Lempitsky,
\newblock ``Domain-adversarial training of neural networks,''
\newblock {\em J. Mach. Learn. Res.}, vol. 17, no. 1, pp. 2030--2096, 2016.

\end{thebibliography}

\end{document}